# A new neural-network-based model for measuring the strength of a pseudorandom binary sequence

Ahmed Alamer, Ben Soh

*Department of Comp. Science and Information Technology*
*School of Engineering and Mathematical*
*Sciences La Trobe University, Victoria,*
*Australia 3086*
*a.alamer@latrobe.edu.au,*
*B.Soh@latrobe.edu.au*

*Abstract*: Maximum order complexity is an important tool for measuring the nonlinearity of a pseudorandom sequence. There is a lack of tools for predicting the strength of a pseudorandom binary sequence in an effective and efficient manner. To this end, this paper proposes a neural-network-based model for measuring the strength of a pseudorandom binary sequence. Using the Shrinking Generator (*SG*) keystream as pseudorandom binary sequences, then calculating the Unique Window Size (UWS) as a representation of Maximum order complexity, we demonstrate that the proposed model provides more accurate and efficient predictions (measurements) than a classical method for predicting the maximum order complexity.

1. **Introduction**

Maximum order complexity is an important tool for measuring the nonlinearity of a pseudorandom sequence. The maximum order complexity of a given sequence is referred to as the function acting as the shortest nonlinear feedback shift register (*FSR*). The *FSR* can generate this sequence [1]. A greater length of the *FSR* is better; hence, this sequence is more resistant to attack and more pseudorandom.

In this regard, to be able to predict the maximum order complexity is important for determining how this sequence has a random appearance [2, 3].

By calculating the unique window size (*UWS*) [4] as a representative of maximum order complexity, as will be explained later, we will be able to establish the measurement of pseudorandomness.

Once the *UWS* has been calculated, we have an overview of the behavior of binary sequence pseudorandomness. Using different degrees of *UWS*, as will be shown, and utilizing a predicting tool will add an advance step to generalize the measurements with the given data. This, in turn, allows us to establish the efficiency of the cipher used in the encryption. This study uses the keystream generated by the shrinking generator as a binary sequence. This is because the study of binary sequence pseudorandomness is the building block for determining the efficiency of ciphers [5].

**1.1. Contribution of this Study**

The main contributions of this study are as follows:

1. Use the *UWS* as a representative maximum order complexity tool for the shrinking generator keystream (binary sequence).

2. Introduce a neural-network-based model for predicting the *UWS*. This will help in evaluating the strength of the cipher in the study, which can be used for similar ciphers with some modifications to the model. furthermore using *NN* models for maximum order complexity at the best of our knowledge is new approach in this regard.

In addition, it can be used for internal cipher components, other than the keystream, that also generate binary sequences, and hence, will help in evaluating the strength of cipher components.

Furthermore, this will inspire other applications of *NN* models in cryptography, especially in encryption, and in security in general, such as in securing communications.

The paper is organized as follows: section 1 provides the background of the study; section 2 introduces *SG* and *UWS* and also provides a brief overview of *NN*s; section 3 reviews the current literature in the field of *NN*s with a focus on its implementation in security; section 4 discusses the proposed *NN* models and their implementation; section 5 discusses the observations of the models and the results; and the conclusions are drawn in section 6.

## 2. Background

This section reviews necessary information and background to understand the basic structure of a *NN* model. As well as, necessary information and background to understand the basic structure of a *NN* model. By focusing on the information associated with our study of the pseudorandom binary sequence resulting from the calculation of *UWS*, with their significance clarified. We also discuss the *SG* cipher, where we describe its principle, which is the source of the data we obtained. We also discuss the *SG* cipher, where we describe its principle, which is the source of the data we obtained.

### 2.1. Unique Window Size

The *UWS* is a type of maximum order complexity tool that is used in this study. This concept can be defined by assuming that the maximum order complexity of a given sequence $S$ is $m$. Then, the *UWS* for $S$ is $m + 1$. We use the *UWS* as a nonlinear measurement of a binary sequence that illustrates how such a sequence has a high level of security. This is necessary for cryptosystems as we need a pseudorandom sequence for these types of security applications. For more information on the statistical behavior and distributions of *UWS*, please refer to [4].

### 2.1.1. Importance of Unique Window Size

*UWS* is a tool for measuring the strength and randomness of a given sequence, even though we use it for an SG as an example. However, it is applicable for testing any ciphers that can generate a sequence with random appearance.

The prediction model can be helpful for the two *LFSRs* choices for shrinking generating and where the choice of the *LFSRs* combinations will result in a small or large *UWS*, so the user can choose the *LFSRs* pair that resulted in a large *UWS*.

By analyzing this, we can detect possible attacks on the given combinations. This is a good direction for further investigation as the sequence with a low UWS can be easily simulated. Furthermore, it can be vulnerable to certain attacks such as a correlation attack [6] and divide and conquer attack [7].

Researchers can adapt *NN* for different purposes needed for prediction of, for example, other randomness measures. The prediction of *UWS* is an example that was implemented in the security field [2].

## 2.2. Neural Networks

*NN*s are a useful cryptographic tool that can assess an algorithm to generate a binary sequence, making it useful in certain industries, such as security [8]. The vital role of *NN* in pattern recognition is discussed in [9].

*NN*s have previously been used for generating secret keys for given ciphers. However, to the best of our knowledge, there are very few studies on the use of *NN* as a tool to measure the pseudorandomness of a binary sequence [10].

A *NN* simulates the behavior of neurons in the brain. It can deal with complex processes, and an algorithm can be developed that can be used to model and predict complex data and compare this data with the original input data. For example, to measure the accuracy of a model, an algorithm is trained and modified by a number of steps until the optimal model is obtained. This can be done by selecting a powerful algorithm for learning and providing suitable and correct inputs [11].

*NNs* scan be applied to cancer prediction [12], weather forecasting [13], image recognition [14], music production [15], and stock market prediction [16], to name a few. They can also be applied in encryption where they can be used to select secret keys, as well as in other cryptographic applications. Predicting the *UWS* is a challenging task as it is calculated using a pseudorandom binary sequence. Therefore, predicting the *UWS* using *NN* will contribute to determining the security of a given binary sequence [17]. The purpose of this study is to evaluate the use of *NN* as a method for evaluating the strength of a pseudorandom sequence. A multi-layer model is used to determine the best prediction model.

## 2.3. Shrinking Generator

*SG* is a cipher that implements two primitive polynomials acting as linear feedback shift registers *(LSFRs)*. To illustrate the concept, assume that $LFSR_a$ is the input *LFSR* and $LFSR_b$ is the controlling *LFSR*. Thus, the binary input from $LFSR_a$ will be controlled by $LFSR_b$ in such a way that if the input bit of $LFSR_b$ is $1_s$, then the bits from $LFSR_a$ will be selected. Otherwise, it will not appear in the keystream sequence. Table 1 lists the selection role for the *LFSR* pairs. In an *SG*, $LFSR_a$ is linear and $LFSR_b$ makes the keystream sequence to be a nonlinear sequence [18]

| $LFSR_a$ | $LFSR_b$ | Keystream bits output |
|---|---|---|
| 1 | 1 | 1 |
| 0 | 1 | 0 |
| 1 | 0 | Not appear |
| 0 | 0 | Not appear |

TABLE 1: *SG* keystream bits selection rule

As *SG* is a lightweight stream cipher, it can be used in small systems with limited power resources, such as in RFID systems. The shrinking concept is important to the study of data management and data selection rules. In addition, an analysis needs to be performed to identify limitations and establish an optimal testing model to be used for a chosen cipher. In an *SG*, the prediction model can illustrate the best combination of *LFSR* pairs that can generate a large *UWS* and prevent combinations with a lower *UWS*. An interesting *SG* variant is a self-shrinking generator (*SSG*) [19], which uses a similar design concept with the difference being that *SSG* uses one *LFSR* and selection and bit input is performed in this *LFSR*.

Jansen [3], in his study of any given maximum order complexity, stated that some sequences have a subsequence with nonlinear complexity. If the occurrence of these subsequences is large, this will eventually weaken the nonlinearity strength, which is not desirable for the randomness needed for the given sequence. Hence, it is not attractive in cryptographic practice. This is important for the sequence to have high nonlinear complexity.

To explain *UWS* in this context, assume we have a sequence S with length *N*. We look for a subsequence of *S* with a sliding window of length *W* (*W*=the number of bits of this subsequence). Then, we choose a sliding window having a length of *n* bits starting from the first bits of the sequence *S* (from the most left). Then, we take the same *m* starting from the second bits and leave out the first bit, then *m* starting from third bit and leave out the second bits, etc. Therefore, if we determine the repetition of states (one or more states appear more than once) by sliding one bit at a time, we can select another sliding window with length *m* + 1. This process is repeated *n* times till every state (subsequence) is unique (without repetition).
At this stage, we chose $W = m + n$ as our *UWS*.

Example 1:

$X^2 + X + 1$, work as $LFSR_A$ acting as input *LFSR* (a primitive polynomial of degree 2, $X^i$ represent the taps position *i* is the tap place in LFSR polynomial, the taps in this example at 2nd and first bits);

$X^5 + X^3 + X^2 + X + 1$ ($LFSR_B$ as control *LFSR*), *UWS* = 15

111101101011101010001101011011110111101111010101

Here, we can see that every state is unique for a sliding window of length $w = 15$. If the length of the sliding window is less than 15, e.g., 14, the subsequence 11011110111101 is repeated twice. If the length of the sliding window is 12 or 13, more states are repeated.
Here, we can see that *UWS* is another important measurement tool of cipher strength as when *UWS* < 15, the states are repeated twice, as in the case with $w = 14$. With brute force calculation, it is possible to find the remaining states [20]. Thus, if we can define *UWS*, we can find subsequences with *UWS* < 15. In addition, finding the correlation between the repeated states and remaining states can lead to a correlation attack, and the correlation between the repeated sequence and *SG LFSRs* pairs can be investigated.
Table 2 lists the number of simulations per *SG* degree to obtain the *UWS*.

| UWS degree | Number of Simulations |
|---|---|
| 7 | 20 |
| 8 | 24 |
| 9 | 24 |
| 10 | 72 |
| 11 | 208 |
| 12 | 216 |
| 13 | 840 |
| 14 | 1280 |
| 15 | 1280 |
| 16 | 6360 |
| 17 | 13080 |
| 18 | 13896 |
| 19 | 48600 |
| 20 | 70416 |
| 21 | 245628 |

TABLE 2: Number of simulations per *UWS* degree for the *SG*

A long keystream attack is a type of security attack. To perform a long keystream attack, knowledge regarding the length of the keystream is required as this type of attack assumes that the attacker has access to long portions of the keystream, as well as with the need of the *LFSR* to have a large number of taps [21, 22]. *UWS* can be used as a measurement tool to evaluate the strength of the sequence.

3. **Related work**

Erdmann et al. introduced in [2] an approximate maximal order complexity distribution for a pseudorandom sequence and compared the distributions obtained in their simulations with actual distribution to develop statistical tests. They emphasized the importance of finding such a distribution for measuring the strength of a pseudorandom sequence.

Sun et al. [23] investigated the maximum order complexity for a given sequence with the aim of determining the relationship between the maximum order complexity and the number of periodic sequences that can be generated with maximum order complexity to use that as an indicator of the nonlinearity strength of the sequence. However, in our approach, not only is the *UWS* generated by the cipher (*SG* in our case), which is based on the maximum order complexity, but the *UWS* is also predicted with high accuracy by implementing the *NN* model.

Sun et al. [24] also found that the maximum order complexity is a better complexity measure compared with expansion complexity, which was applied to the Thue−Morse sequence and the Rudin−Shapiro sequence.

Kenzel et al. [25] were inspired by public cryptography methods using a public key exchange, which was introduced by Diffie and Hellmann in 1976. Using these methods, the public key can be shared in an insecure channel and is accessible to the public. A discrete logarithm was used for the public key, making it difficult for devices with limited computation power to handle, especially in cases where a large number of public keys was chosen. thus investigated interactive *NN*s to investigate secret key exchange through a public channel. In addition, Godhavari et al. [26] implemented an interactive *NN* to generate the secret key over a public channel using the DES algorithm.

Meidl et al. [27] demonstrated multiple sequences over a given finite field by analyzing two groups of nonlinear complexity with interactive differences between them. In addition, to create a probability of joint nonlinear complexity over a given fixed finite field, they introduced joint probability based on their parameters with special conditions.

To summarize, it is important to have a clear understanding of the nonlinear complexity and behavior of a binary sequence and to analyze the optimal methods to evaluate it. The prediction of *UWS* is an important direction in this regard.

With respect to studies on the applications of *NN* to security, Allam et al. [28] presented a binary tree algorithm that can be implemented by a mutual learning process to ensure secure communication through exchange of keys between groups in public cryptography methods. They showed that the complexity of their algorithm is logarithmic and depends on the number of parties they synchronized together. In addition, they introduced the possibility of using *NN*s for secure key sharing. Furthermore, Fan et al. [29] explained how *NN* is a powerful prediction tool by using an *NN* to find the correlation between the pseudorandom data (especially binary data) and used this to predict the bits based on the neighboring bits. In their study, they dealt with multiple pseudorandom sequences and they showed how *NN* can learn from a given dataset to predict the outcomes in testing data.

Introducing a new method for predicting the maximum order complexity (*UWS* in our case) for the pseudorandomness of a binary sequence by *NNs* models with high accuracy is new and important direction for evaluating the cryptosystems used. as well as predicting the pseudorandomness for the binary sequences in general.

**4. Proposed neural-network-based prediction for pseudorandom sequences**

We ran the calculation of *UWS* for the SG keystream (example 1 shows how *UWS* was calculated) on *EC2*, which is a service provided by Amazon Web Services for cloud computing, such as UWS with degree 20 (*UWS20*). We used a cloud server (*EC2*) as it enables us to simultaneously run multiple data files, thus reducing the time required for calculating UWS. A large number of simulations are required for computing *UWS*, for example, *UWS20* has a data size 70416, which requires 70416 simulations.

We used Linux on *EC2* for calculating *UWS*. A similar procedure was employed for *UWS* with degree 21 as well as for all *UWS* with degrees ranging from 7 to 19, as discussed later. Let the primitive polynomial $X^{17}+X^5+X^3+X^2+1$ as $LFSR_A$ and $X^3+X^2+1$ is $LFSR_B$.

Then the SG degree is 20 by adding the left most power from the LFSRs(17+3), Similarly, when the highest powers of the other combinations of $LFSR_A$ and $LFSR_B$ are added, they sum to 20. Then, we obtain all 70416 possible *UWS* for *SG* with degree 20(*UWS*20). The independent variables within the prediction model are input degree, input weight, control degree, and control weight. The dependent variable is the *UWS*, as shown in the following example:

Example 2:

Let us take the following example: $LFSR_A$(input *LFSR*) as $(X^{13}+X^9+X^8+X^6+X^4+X^2+1)$, and $LFSR_B$(control *LFSR*) as $(X^7 + X^6 + X^4 + X^2 + 1)$, then the input $LFSR_A$ degree = 17, $LFSR_A$ weight =7, control $LFSR_B$ degree =7 and $LFSR_B$ weight =5,in this case *UWS* = 39.

For the sake of clarity, let us assume that we have two *SG* polynomial combination samples for *UWS*20, as shown in Table 3:

| Input Degree | Input Weight | Control Degree | Control Weight | UWS20 |
|---|---|---|---|---|
| 3 | 3 | 17 | 3 | 42 |
| 3 | 3 | 17 | 7 | 39 |
| 7 | 5 | 13 | 7 | 44 |
| 11 | 7 | 9 | 7 | 42 |
| 13 | 7 | 7 | 5 | 39 |
| 17 | 5 | 3 | 3 | 36 |
| 9 | 7 | 11 | 7 | 45 |
| 17 | 5 | 3 | 3 | 37 |

TABLE 3: *UWS*20 sample for illustration

To analyze the results, the *NN* model was implemented using Keras [30] as a tool with tensorflow [31] as a backend. The *UWS* with degree 20 (*UWS*20) was computed (see example 1) for the keystream sequences generated by the *SG* cipher, and all possible combinations of *LFSRs* of the *SG* were obtained. There were 70416 possible combinations, and we calculated the *UWS20* for all the combinations with the *NN* model using a ReLU activation function for the outer layer, and the ReLU activation function for the hidden layers (Table 4: Model summary). We obtained a prediction with high accuracy as shown in Table 5 in addition to *UWS*21 for comparison.

**4.1 Implementation**

We use multiple functions of the Keras library in Python to build our model. These features help us optimize the performance of the model and training time. These functions are listed below:

**1. Earlystopping:** This function stops the training of the model if the accuracy of the model does not show any improvement after a pre-defined number of epochs.

**2. Sequential:** This function helps define the model type of *NN*. A *NN* can have multiple layers, and this function helps the model learn each layer.

**3. Dense:** This function adds a fully connected layer to the *NN* model and makes the neurons in the layer to be connected to the neurons in the next layer.

**4. Optimizer.Adagrad:** This function defines the optimizer for training the model.

**5. Compile:** This function defines the parameters for the model.

**6. Fit:** This function trains the model by updating the weights in the model.

**4.2. Mathematical description of NN Model**

Let us consider a matrix *MN*, where *M* is the number of *UWS* simulations and *N* is the number of features, which include input degree, control degree, input weight, and control weight. We have *D* number of layers in the NN, and these layers include the input layer, output layer, and *G* number of hidden layers, where each hidden layer is for learning the input data. A greater number of layers in the NN will help in learning more complex data. The relationship between the input layer and first hidden layer is represented by matrix multiplication. However, if we start from first principles, we can assume there is just one example, and it is a column vector of size $N*1$.

If $N = 3$ and we assume that the first hidden layer had $H_1 = 4$, each layer will be fully connected so that every neuron in the hidden layer is influenced by every neuron in the input. layer. Let $H_{1j}$ be the $j^{th}$ neuron of the hidden layer $j$. Consequently, the output is calculated as the dot product of the input layer neurons and the weights connected to the $j^{th}$ neuron of the hidden layer. Thus, let $w_{1ij}$ be the weight from input neuron $i$ to hidden layer neuron $j$, $x_{ij}$ is the example feature (independent variable); thus,

$$H_{1ij} = \sum_{i=1}^{N} W_{1ij} . X_{ij}$$, and $W$ denotes the weights matrix

We can represent this in the multiplication matrix, where each row denotes the result of the hidden layer for each example:

$$H_1 = X \times W_1^T$$

$W_1$ is a matrix that is $N \times H_1$, where $N$ denotes the number of features for an input example and $H_1$ denotes the number of neurons for the hidden layer. We transpose $X$ because each example is in a column for the mathematics to work, but the original matrix $X$ is a row matrix. For *NN*s, each layer is represented as a matrix of weights. For our particular case with input data (for *UWS20*), the first hidden layer has 100 neurons, the second layer has 50 neurons, third layer has 20 neurons, the fourth layer has 10 neurons, and the output layer has 1 neuron.

After this operation, to model the nonlinearities in the data, this goes through an activation function. In our case, we used the *rectifier linear unit (ReLU)* activation function [32] a single value $x$ it is defined as follows:

$$R(x) = max(0,x)$$

$$R(x) = x \text{ when } x > 1, R(x) = 0 \text{ (otherwise)}$$

note our datasets are positive integers; hence, using ReLu presents advantages as our data do not have negative values.

Therefore, the *ReLU* activation function works on a per element basis. Thus, we apply this to each output value of $H_1$. Thus,

$H'_1 = \sigma(H_1)$, σ is the activation function(R)

For the second output layer, we repeat the same operation.
$$H_2 = H'_1 \times W_2^T$$

$H_2$ contains the outputs for the second hidden layer, considering the activation function.
$$H'_2 = \sigma(H_2)$$

The same procedure is used for the third layer Finally, we have the output layer:
$$H_0 = H'_3 \times W_4^T$$

In summary, the output layer is set to the total number of expected outputs from our model will be a Matrix with single column contains *UWS* (*UWS20* has 70416 elements). In this case, we are estimating a single quantity so that there is just one output neuron for that layer. The number of input neurons in the input layer is the same as the number of features.

The number of neurons in the hidden layer is usually the average of the input and output neurons. Thus, for our case, the number of neurons is $\frac{4+1}{2} = 2.5$; thus, we round this off to 3 neurons. However, the performance was poor. Therefore, we tried to increase the number of layers by 1 so that there are 4 hidden layers. The first hidden layer has 100 neurons because of a lack of available features. We want each neuron to provide a good representation for the lack of available features. The second layer has 50 neurons, the third layer has 20 neurons, and the last hidden layer has 10 neurons and this layer is known as a "bottleneck" layer. Thus, once we learn the complex representations, we force the network to remember the best things about representation.

| Layer (type) | Output Shape | Param # |
|---|---|---|
| dense_1 (Dense) | (None, 100) | 500 |
| dense_2 (Dense) | (None, 50) | 5050 |
| dense_3 (Dense) | (None, 20) | 1020 |
| dense_4 (Dense) | (None, 10) | 210 |
| dense_5 (Dense) | (None, 1) | 11 |

Total params: 6,791,Trainable params: 6,791

TABLE 4: Model summary from the output of Python code on Keras

**4.2.1. Determining optimal weight matrices.**

To determine the optimal weight matrices, we have to define a cost function that determines the penalty in case the predicted outputs are dissimilar, given the *NN* structure with the true output values.

In our case, we defined the mean squared error or the sum of the squared errors. The output of $H_4$ will be a matrix of $M \times 1$, where $M$ is the total number of examples and there is just one column for the predicting output (UWS).

To calculate the cost function, let $y_i$ be the true output value for $x_i$, and $H_{4i}$ is the predicted output for *i*. Therefore,

$$C = \tfrac{1}{2M} \sum_{i=1}^{M} (H_4 i - y_i)^2$$

Thus, we multiply by $\tfrac{1}{M}$ because we want to find the mean, and we additionally $\tfrac{1}{M}$ divide by 2 when optimizing (finding) the derivative.

Given a batch of training examples, Keras will optimize the weights by minimizing the cost function. These sets of weights can be found analytically by using pseudo inverse, but this will consume a considerable amount of memory, particularly if there are a lot of examples that require a lot of computational time.

The alternative is to use a method called backpropagation that systematically determines the weights, one layer at a time. This leads to a method called gradient descent, which is an iterative method to determine the minimum of a function. The procedure can be described as follows:
1. Set the weights of each hidden layer to random.
2. Decide on a batch of examples to feed into the network. The batch size directly affects the weights. The larger the batch size, the more accurate the weights. This will hopefully lead to faster convergence. Smaller batch sizes are sometimes used because the network can be very deep (large number of layers) and the number of examples can also be high.
3. Do a forward pass, which involves taking examples and finding the outputs of each layer as well as the output layer.
4. Compute the backward pass (backpropagation) where the inputs from step #2, as well as the current weights, are used to update and find the gradients with respect to the weights
5. Compute an update for gradient descent.
6. Repeat steps 2 to 5 until the network converges (The cost function should decrease over time and we can stop if it does not change considerably or if we impose a maximum number of iterations / epochs). One epoch amounts to the number of batches of examples we need to go through in the entire dataset once. For example, if there are 1024 examples and the batch size was 32, a total of 32 iterations would need to be performed for 1 epoch (because 32×32 = 1024).

In our case, the batch size is 8, and the number of epochs is 100. The number of examples is 70416. Therefore, the number of iterations per epoch is $\tfrac{70416}{100}$ = 8802. As we need to consider all examples, we select 8802 examples per epoch. The total number of iterations will be 8802×100 = 880200.

**4.3. Scaling the dataset for training**

The dataset is scaled so that the model can learn faster

The MinMax scaling method scales the feature set between values [0,1] by using the following formula:

X_scaled = ($X_i$ - X_min)/(X_max - X_min) for each *i*

- *i* = The element number on which the transformation is occurring
- $X_i$ = The value of the *i*th element for the selected feature
- X_max = The maximum value (among the dataset) of the feature
- X_min = The minimum value (among the dataset) of the feature
- X_scaled = The transformed value (among the dataset) of the feature

The sklearn library provides a function MinMaxScaler() for this purpose.

### 4.4. Splitting the dataset for training & testing

The datasets are generally split for training and testing. Generally, 80 sets are used for training, and 20 sets are used for testing. The validation split ratio is used for dividing the datasets.

X_train, X_test, y_train, y_test = train_test_split(X, y, test_size=0.20, random_state=123)

### 4.5. Estimation of performance of proposed model

We used a custom-built accuracy calculating measure to evaluate the performance of the model. The accuracy of the model is 91.90% as per the calculations (for *UWS*21) see Table 5.

### 4.5.1 Performance estimation code:

1-Obtaining the predicted values for the test data
 y_pred = model.predict(X_test)

2-Unscaling the predictions: by unscaling the original and the predicted output (using the range and the min values from the original dataset)
y_predscaled = [int(i*rangel + rangebot) for i in y_pred]

3-Unscaling the test: obtaining the deviation values from the original dataset.
y_testscaled = [float(i*rangel + rangebot) for i in y_test]
dev = []
for i in range(len(y_pred)):
dev.append(abs((y_predscaled[i]-y_testscaled[i]))/y_testscaled[i])

4-Obtaining the accuracy number: the efficiency of the model is measured using the deviation = 1 - sum(absolute_error)/sum(true_value)

1 - sum(dev)/len(dev)

The average deviation of the actual value from the predicted value is 5.6% (across the test dataset)

The accuracy of the proposed prediction model is high when predicting the *UWS*, which is an important tool for measuring the pseudorandomness of a given binary sequence. Hence, this model will help designers in choosing the best cipher components, which make attacks complex and resource consuming [33, 34].

*UWS* is important for understanding the binary sequence, if it is predictable, and is another tool among other method recently proposed for the same purpose [35].

### 5. Prediction results

The prediction accuracy for *UWS* with degree 20 was 94.3 % of the test data and that for UWS with degree 21 was 91.9%. Table (5) summarizes the findings for the 2 *NN* models with degrees 20 and 21.

|  | *UWS*20 | *UWS*21 |
| --- | --- | --- |

|                  | Model        | Model        |
|------------------|--------------|--------------|
| Number of Layers | 4            | 4            |
| Number of nodes  | 100,50,20,10 | 100,50,20,10 |
| Learning rate    | 0.0001       | 0.0001       |
| MSE              | 0.0088       | 0.0064       |
| training set     | 56333        | 196502       |
| Accuracy         | 94.30%       | 91.90%       |

TABLE 5: *UWS*20 and *UWS*21 Model comparison

From Table 5, we can see that the mean squared error of the model as the cost function is 0.0.0064 for *UWS21* (for example) on the MinMax scaled dataset (see 5.1). The MSE is quite low and suggests that the model is very good, as listed in Table 5, in addition to the results in Table 5, we combine all datasets from UWS7 to UWS19 in one dataset, and using one NN model for the Joint datasets (see Table 2 for the number of samples for each UWS degree), to see the difference once we train one model contain different UWS degrees, with combined dataset of size 85900, and training 80% of which is used for training (68720), hence the accuracy is 93.7%, with a learning rate 0.001 and MSE = 0.0028 which is very low and the number of hidden layers is the same as in the UWS20 and UWS21 models. The reason behind using one model is that we want to investigate if the model is accurate when using *UWS* with degrees varying from 7 to 19 so as to determine whether the model is still valid compared with the results in Table 5 (using separate models for *UWS*20 and *UWS*21).

**5.1. Model Features Influence:**

There are multiple ways of calculating the importance of features. It is important to understand the most independent variables that influence model accuracy. By evaluating the properties of the futures after the model fitting, as we want to determine which among them has most impact in the predicting results. Thus, we would not change the model or change the predictions we will obtain for a given value of feature. Instead, we want to determine how randomly shuffling a single column of validation data, leaving the target and all other columns in place, would affect the accuracy of predictions. Randomly reordering a single column should result in less accurate predictions, as the resulting data no longer correspond to anything observed in the real world. Model accuracy especially suffers if we shuffle a column that the model relied on heavily for predictions. In this case, shuffling control degree would result in inaccurate predictions. If we shuffled control weight instead, the resulting predictions would not be as inaccurate.

As per the analysis, we are confident that control degree is the most important feature in deciding the value of the dependent variable (the influence value is the highest: 0.0077) followed by input degree, input weight, control weights for *UWS* with degree 21 as listed in Table 6.

in the following the implemented code for the evaluation of the model features influence on the prediction results:

```
{
perm=PermutationImportance(model, random_state=1,scoring="neg_mean_squared_error").fit(X_test, y_test)
eli5.show_weights(perm,feature_names=dataset.columns.tolist()[:4])
}
```

Therefore, knowledge on the important inputs (independent variables) and finding the most important variable will have a greater impact on the forecasting result, It will also provide a clear idea of the sources of strength and weakness in the cryptosystem, which will also reflect on our ability to develop it. In addition to this, with further research and investigation in this direction, it may be possible to develop a mathematical relationship to analyze and represent how to obtain a more random sequence. Moreover, our approach can be applied to any source that can produce a random sequence by knowing the strengths and weaknesses based on the components of the source by calculating the *UWS*, because the bigger the size the better. Thus, we can test and select the best encryption systems.

| Feature | Influence |
|---|---|
| Control Degree | $0.0371 \pm 0.0002$ |
| Input Degree | $0.0293 \pm 0.0004$ |
| Input Weight | $0.0066 \pm 0.0001$ |
| Control Weight | $0.0045 \pm 0.0001$ |

TABLE 6: *UWS21*: Influence of features in prediction

**5.2. Overall Observation**

The *NN* in our research has clearly demonstrated that it is an effective tool for prediction, which is reflected in its ability to measure the effectiveness of any cryptosystem, based on a pseudorandom binary sequence as the pseudorandomness measurements for the binary sequences which can be produced by this system. Thus, it serves as a measuring tool as well as a system that can be developed with more research and investigation to advance the encryption systems used. Therefore, *NN*s help in analyzing the strength of the security system used based on the sequence it produces. By measuring the model's ability to predict, the strength of the cryptographic system can be determined, and thus, the system can be developed by designers to be more resistant to prediction. It also helps users in selecting the best encryption systems and the most effective to resist attacks, not for obtaining that the applications of the *NN* are being expanded to different fields, especially in information protection systems. This will assist in their future development.

Calculating and measuring *UWS* predictability will help in the design of ciphers with larger complexities with respect to nonlinearities, which is a necessary condition for sequence pseudorandomness. as the larger the *UWS* is the better; the complexities of attacks will also be greater [33] in the future, indicating that ciphers using nonlinear feedback shift registers should be investigated in depth owing to their increasing usage. Further, the sequences generated by these ciphers need to have high values for maximum-order complexity [34]. As the NN models can generate results with high speed, it would be advantageous to investigate the subsequences of a given binary sequence, which is an interesting direction for future research on linear complexity profiles [36], as well as to evaluate NN model applicability as large correlation measures of automatic sequences [37].

**6. Conclusion**

*NN*s have been proven as good prediction tools for *UWS*, which can help in investigating the strengths of pseudorandom binary sequences and their generators. Our approach is applicable to any sequence generator, such as a cipher, which can help designers and users who require such sequences or ciphers for specific applications. *NN*s are very efficient prediction tools even when the data have random features. In the specific case investigated in our work, the data are discrete and cannot be represented by a continuous statistical distribution, making it difficult to implement basic and classical statistical analysis methods efficiently and with reasonable accuracy. Hence, the *NN* is an efficient and powerful alternative, as shown by our findings. Some future directions for this work that are worth investigating will further strengthen and generalize this method for obtaining the *UWS* on the *SG* keystream are as follows:

1) Applying the proposed model to the keystreams of other ciphers, for example, other stream ciphers (e.g., Trivium, Mickey) and block ciphers such as *AES*, *DES*.

2) Applying the model to binary sequences generated by internal cipher components, such as *FSR* and *LFSR*. In the *SG* case, for example, *LFSRs* will generate binary sequences. Calculating the *UWS*s for these sequences and applying the model would be of interest for evaluating the strengths of the internal components.

3) An additional direction for research (not related to security) could be facial or image recognition. By converting the input images and features to binary sequences and applying the proposed approach, with some modifications to the prediction model, an alternative tool for pattern recognition can be realized as it has a number of important applications.

**Acknowledgement**

We would like to thank "Lito P. Cruz" for the valuable Neural Network comments, and we would like to thank Editage (www.editage.com) for English language editing.